\documentclass[aps,amsmath,showpacs,amsfonts,10pt]{revtex4}
\usepackage{epsfig,graphicx}
\usepackage[english]{babel}
\usepackage{amsfonts}
\usepackage{amsmath}
\usepackage{latexsym}
\usepackage{graphics,bm}
\usepackage{natbib}
\usepackage{dcolumn}
\usepackage{bm}
\usepackage{rotating}

\begin{document}

\title{Collapse and revival of oscillations in a parametrically excited Bose-Einstein condensate in combined harmonic and optical lattice trap }

\author{Priyanka Verma$^{1}$, Aranya B.\ Bhattacherjee$^{2}$ and ManMohan$^{1}$}

\affiliation{$^{1}$Department of Physics and Astrophysics, University of Delhi, Delhi-110007, India} \affiliation{$^{2}$Department of Physics, ARSD College, University of Delhi (South Campus), New Delhi-110021, India}

\begin{abstract}
In this work, we study parametric resonances in an elongated cigar-shaped BEC in a combined harmonic trap and a time dependent optical lattice by using numerical and analytical techniques. We show that there exists a relative competition between the harmonic trap which tries to spatially localize the BEC and the time varying optical lattice which tries to delocalize the BEC. This competition gives rise to parametric resonances (collapse and revival of the oscillations of the BEC width). Parametric resonances disappear when one of the competing factors i.e strength of harmonic trap or the strength of optical lattice dominates. Parametric instabilities (exponential growth of Bogoliubov modes) arise for large variations in the strength of the optical lattice.
\end{abstract}


\maketitle

\section{Introduction}

A periodic variation of a system parameter leads to the exponential growth of certain modes of the system. This phenomenon is known as parametric resonance and occurs in a wide variety of systems such as classical oscillators, nonlinear optics, system governed by non-linear Schr\"{o}dinger equations and in Hamiltonian chaotic systems. Parametric excitations in the form of Faraday waves, were observed in a cigar-shaped Bose-Einstein condensate (BEC) by periodically modulating the radial trap frequency \citep{Engels07}. The modulation of the radial trap frequency leads to a periodic modulation of the density of the cloud in time, which in turn leads to a periodic change in the nonlinear interactions. This leads to the parametric excitation of longitudinal sound-like waves (Faraday waves) in the direction of weak confinement. From the theoretical side, parametric excitations have already been investigated by numerous authors \citep{castin,kagan,kevrekidis,adhikari,abdul1,abdul2,stalinus02,stalinus04,nicol,bhattacherjee}. A BEC in an optical lattice potential exhibits parametric excitations when the depth of the optical lattice potential is periodically modulated in time. These parametric excitations correspond to an exponential growth of the population of counter-propagating Bogoliubov excitations \citep{tozzo1, tozzo2}. When the frequency of oscillation of the scattering length is an even multiple of the radial or axial natural oscillation  frequency, respectively, the radial or axial oscillation of the condensate exhibits resonance \citep{adhikari}. Parametric resonances in a BEC under periodic oscillations of the position of a quadratic plus quartic trap have also been predicted \citep{abdul2}. Experimentally, St\"{o}ferle et al. \citep{stoferle, schori} measured the excitation spectrum of the condensate in the superfluid regime by time modulating the optical lattice depth. The width of the expanding cloud after it is released from the trap is taken as a measure of the excitation energy. The observed resonances in these experiments were explained by the generation of parametric amplification of Bogoliubov states and the subsequent nonlinear dynamics, which leads to the broadening of the momentum distribution \citep{tozzo1, tozzo2}.

In this work, we study parametric resonances in an elongated cigar-shaped BEC in a combined harmonic trap and a time dependent optical lattice by using numerical and analytical techniques. In order to distinguish the present work from parametric resonances studied earlier, we always consider the case when the frequency of oscillation of the optical lattice depth is not equal to the axial or radial natural oscillation frequency. We show that there exists a relative competition between the harmonic trap which tries to spatially localize the BEC and the time varying optical lattice which tries to delocalize the BEC. This competition gives rise to parametric resonances (collapse and revival in the oscillations of the BEC width). Parametric resonances disappear when one of the competing factors i.e strength of harmonic trap or the strength of optical lattice dominates. On one hand, parametric instabilities (exponential growth of Bogoliubov modes) arise for large variations in the strength of the optical lattice and on the other hand, stable oscillations of the condensate width occurs when the harmonic trap strength dominates. We employ a variational technique \citep{victor, juan, mark} to derive analytical predictions for the parametric resonances. The analytical results qualitatively agree with the numerical results.

\section{MODEL}

We consider an elongated cigar shaped Bose-Einstein condensate(BEC) of $N$ two-level $^{87} Rb$ atoms in the $|F=1>$ state with mass $m$ and frequency $\omega_{a}$ of the $|F=1>\rightarrow |F'=2>$ transition of the $D_{2}$ line of $^{87} Rb$ in combined harmonic and optical lattice potential. In order to create an elongated BEC, the frequency of the harmonic trap along the transverse direction should be much larger than one in the axial (along the direction of the optical lattice) direction. This system is described by the standard mean-field model for the BEC, the GP equation in 3D which is

\begin{eqnarray}\label{gp_1}
 i \hbar \frac{\partial\psi({\bf r},t)}{\partial t}&=& [\ -\frac{\  \hbar^2}{2m}\nabla^2
 +V  + NU_{0}\vert\psi({\bf r},t)\vert^2 \ ]\ \psi({\bf r},t)
\end{eqnarray}

where $ \psi({\bf r},t) $ is the condensate wavefunction, m is the mass of an atom, N represents the total number of the atoms in the condensate, $a$ is the atomic scattering length which can be tuned by the Feshbach resonance technique, $g=4\pi\hbar^{2}a/m$ is the three dimensional inter-atomic interaction. The normalization condition of the order parameter is $ \int \vert \psi({\bf r},t) \vert^{2} dr = 1$. The confining potential for the BEC is given by

\begin{eqnarray}
V(x,y,z,t) \ = \ \frac{\gamma}{2} m(\omega_{x}^{2} x^{2}+\omega_{y}^{2} y^{2}+\omega_{z}^{2} z^{2} ) \ +\ V^{OL}_{1D}
\end{eqnarray}

where $\omega_{x}=\nu \omega,\omega_{y}=\kappa \omega,\omega_{z} = \lambda \omega$ represents the trapping frequencies in the respective directions. We have assumed that harmonic potential in the x-direction is superimposed by a 1D OL potential of the form

\begin{eqnarray}
 V_{1D}^{OL} = V  sin^{2}(\frac{2\pi x}{\lambda_{l}}) (1 + \alpha  sin(\Omega t))
\end{eqnarray}

where V is the amplitude of the OL potential. $K =\frac{2\pi}{\lambda_{l}}$ and $\lambda_{l}$ is the wavelength of the laser beam used to create 1D OL potential. $\gamma$ is the parameter representing the strength of harmonic confinement such that $0<\gamma<1$. In our all calculations $\gamma \cong 0.001$ shows very weak and $\gamma=1$ shows very strong harmonic confinement. $\Omega$ is the frequency of external perturbation at any time t in the time dependent 1D OL potential with the parameter $\alpha < 1$.
We are interested in studying the effect of time dependent potential on the dynamics of the cigar-shaped BEC. Taking the y,z directions to be the transverse and x as longitudinal we assume the confinement in the transverse directions to be sufficiently strong $(\omega_{z},\omega_{y}>>\omega_{x})$  so that the transverse degrees of freedom $(y,z)$ freeze out and the condensate motion gets confined to the longitudinal direction only and the BEC is quasi 1D.

To solve the above problem numerically, we first make the Eqn.\ref{gp_1}dimensionless through a set of linear transformations
$ t\rightarrow \frac{\tau}{\omega} \ , \ x\rightarrow Xl \ ,\ y\rightarrow Yl \ , \ z\rightarrow Zl \ ,\ l=\sqrt{\frac{\hbar}{m\omega}} \ ,\ \psi(x,y,z,t)=\frac{\phi(X,Y,Z,\tau)}{\sqrt{l^{3}}}$ which we will use throughout the paper. After these transformations Eqn. \ref{gp_1} becomes

\begin{eqnarray}
i\ \frac{\partial\phi(X,Y,Z,\tau)}{\partial \tau}&=&\Big[\frac{-1}{2}(\frac{\partial^{2}}{\partial X^{2}}+\frac{\partial^{2}}{\partial Y^{2}}+\frac{\partial^{2}}{\partial Z^{2}})+\frac{\gamma}{2}(\nu^{2}X^{2}+\kappa^{2}Y^{2}+\lambda^{2}Z^{2}) \nonumber \\
&+&\frac{V_{1D}^{OL}}{\hbar\omega}+\eta_{3D}\vert \phi(X,Y,Z,\tau)\vert^{2}\Big]\phi(X,Y,Z,\tau)
\end{eqnarray}

where $\eta_{3D}$ is the non-linearity in the GP equation. To reduce the above Eqn to quasi-1D form it is assumed that the system remains confined to the ground state in the transverse direction. By integrating the above Eqn with respect to the ground state transverse wavefunctions using the method of separation of variables,setting $\nu$=1,we obtain quasi 1D form as ,

\begin{eqnarray}\label{gp_3}
 i\ \frac{\partial\phi(X,\tau)}{\partial \tau}=\Big[\frac{-1}{2}\frac{\partial^{2}}{\partial X^{2}}+(\frac{\gamma}{2}X^{2}+\frac{V_{1D}^{OL}}{\hbar\omega})+\eta_{1D}\vert\phi(X,\tau)\vert^{2}\Big]\phi(X,\tau)
\end{eqnarray}

where $\eta_{1D}=2aN\sqrt{\lambda \kappa}/l$ with the normalization $\int^{-\infty}_{+\infty}\phi(X,\tau) dX =1$ and with $\nu$=1 we assume $l$ to be the harmonic oscillator's length in longitudinal direction.
Also $ \frac{V_{1D}^{OL}}{\hbar\omega}=AT sin^{2}\Big( \frac{2\pi X}{\overline{\lambda_{l}}} \Big )$
where $A=V_{o} \frac{2 \pi^{2}}{{\overline \lambda_{l}^{2}}}$ , $T=(1+\alpha sin(\omega_{r}\tau))$ , $V_{o}=V/E_{R}$ , ${\overline \lambda_{l}}=\lambda_{l} l$ , $\omega_{r}=\Omega/\omega_{x}$, $E_{R}  = \frac{\hbar^{2} K^{2}}{2m}$.

Therefore we can rewrite the above Eqn.\ref{gp_3} as

\begin{eqnarray}\label{gp_4}
 i\ \frac{\partial\phi(X,\tau)}{\partial \tau}=\Big[\frac{-1}{2}\frac{\partial^{2}}{\partial X^{2}}+(\frac{\gamma}{2}X^{2}+AT sin^{2}(\frac{2\pi X}{\overline{\lambda_{l}}}))+\eta_{1D}\vert\phi(X,\tau)\vert^{2}\Big]\phi(X,\tau)
\end{eqnarray}

\begin{figure}[t]

\begin{tabular}{cc}
\includegraphics [scale=0.3,angle=-90] {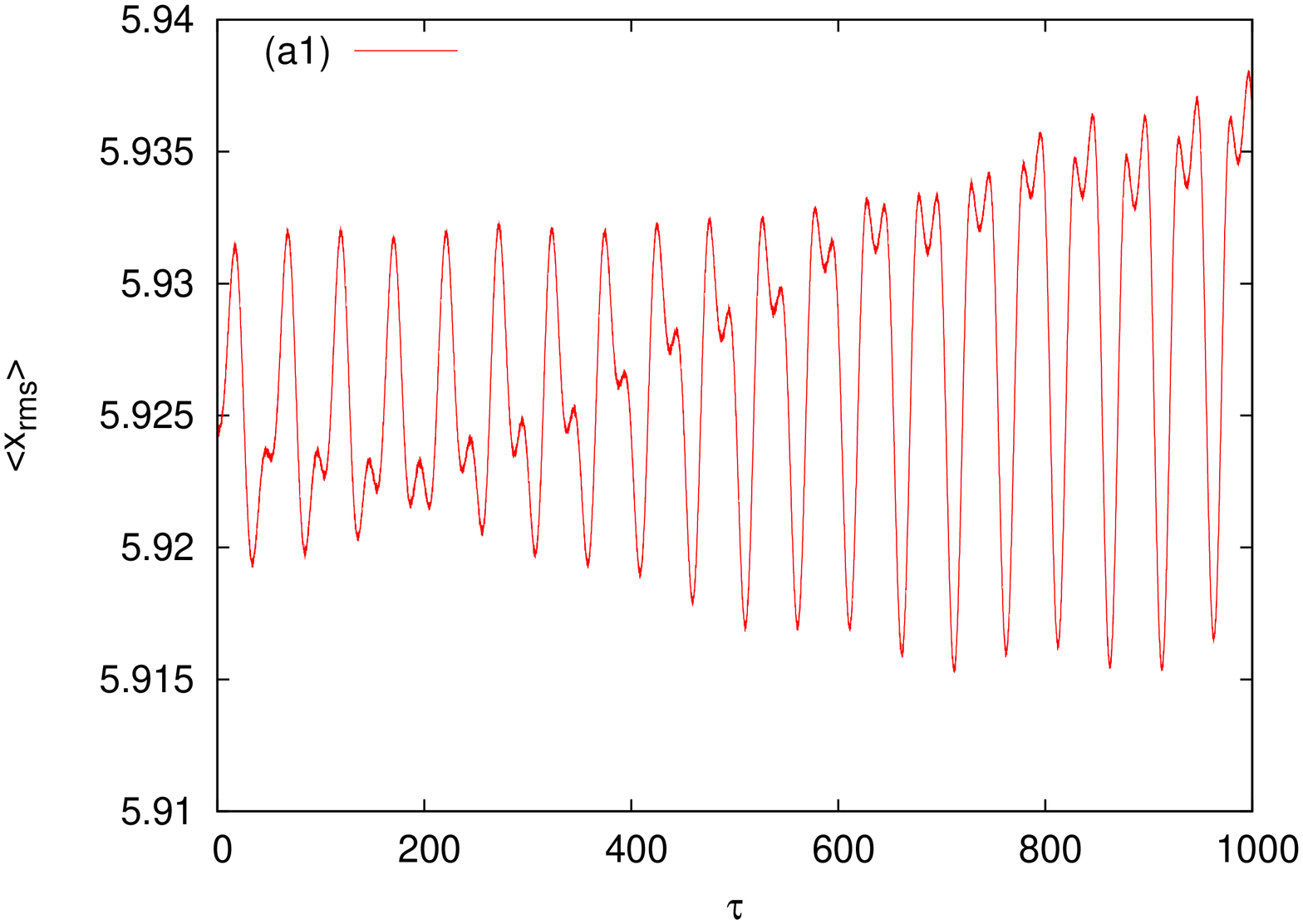}& \includegraphics [scale=0.3,angle=-90] {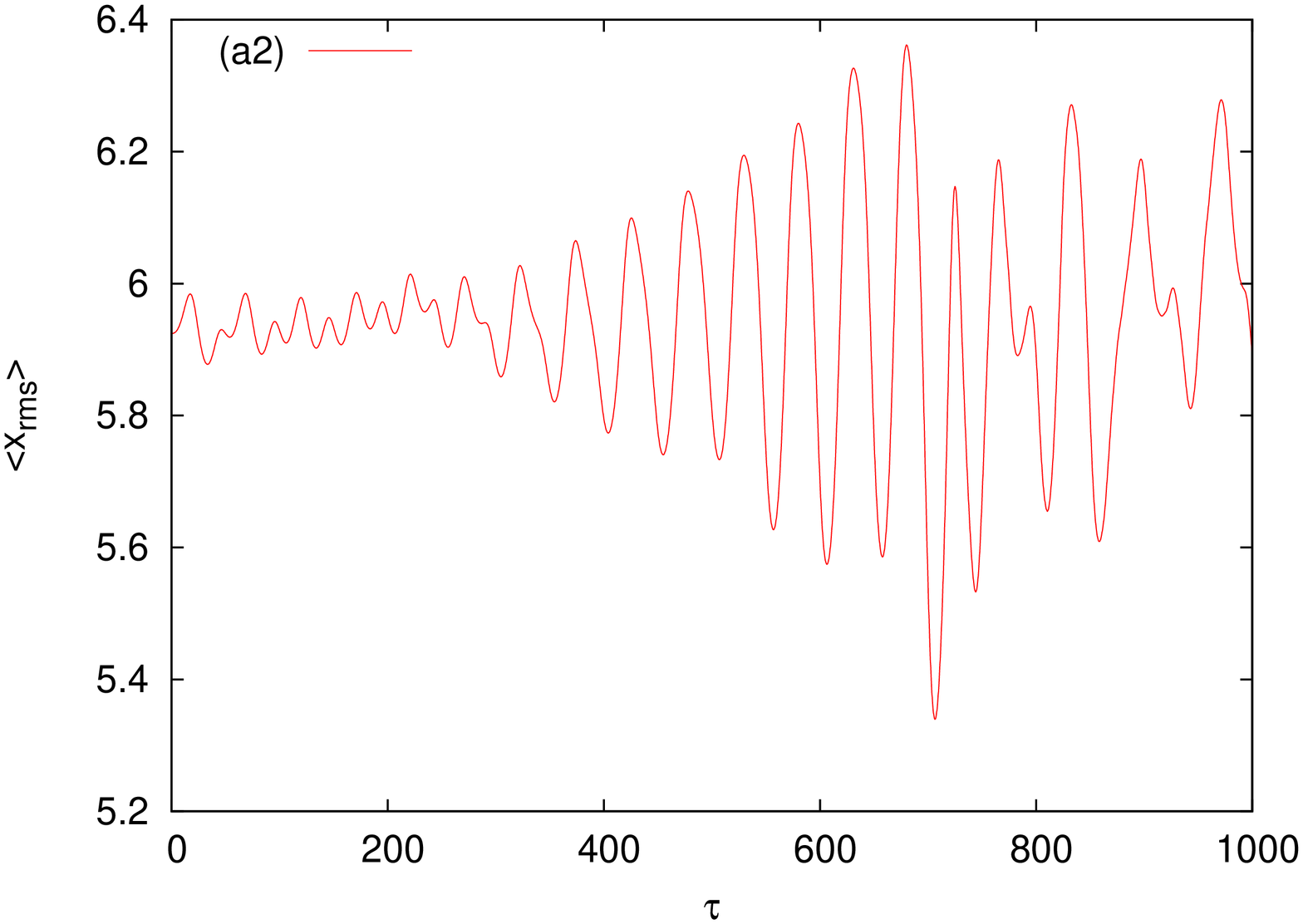}\\
\includegraphics [scale=0.3,angle=-90] {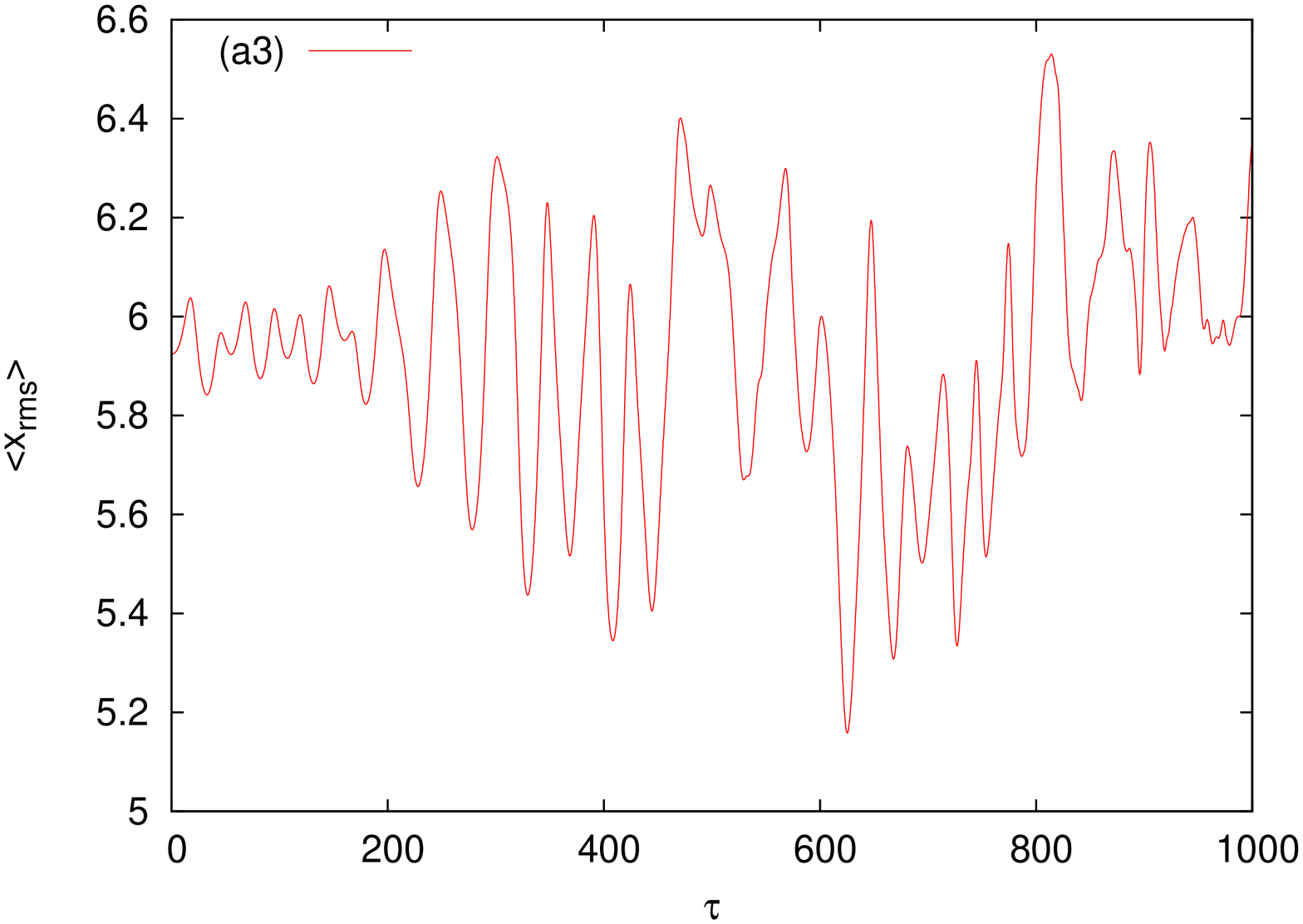} &\includegraphics [scale=0.3,angle=-90] {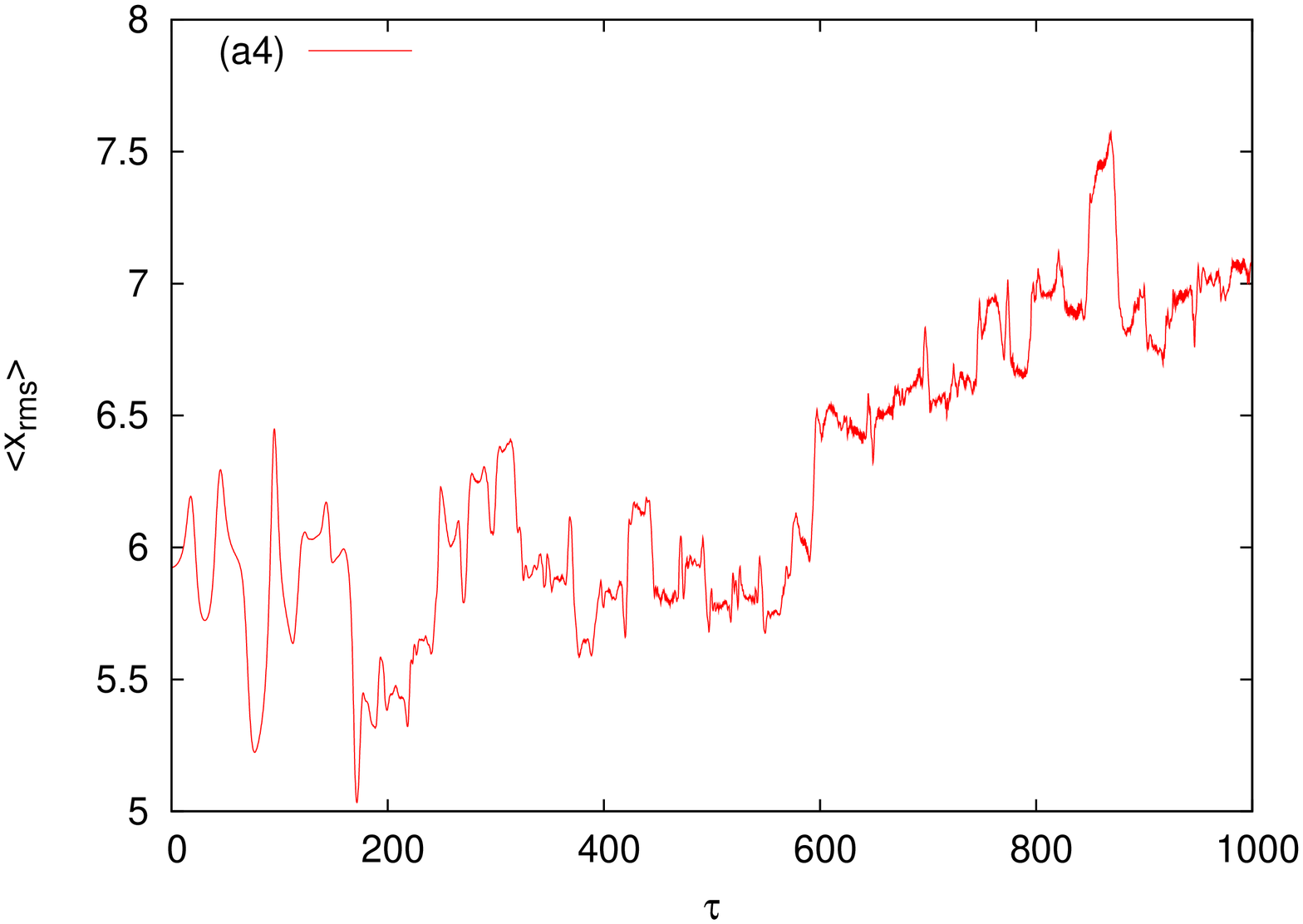}
\end{tabular}

\caption{Parametric resonances and instability in the width of the BEC vs dimensionless time $\tau$ in a combined harmonic trap and a time modulated optical lattice. The parameters are, $\gamma=0.02$, (a1): $\alpha=0.01$, (a2): $\alpha=0.1$,(a3): $\alpha=0.2$,(a4): $\alpha=0.5$.  }

\label{figure1}
\end{figure}

\begin{figure}[t]

\begin{tabular}{cc}
\includegraphics [scale=0.3,angle=-90] {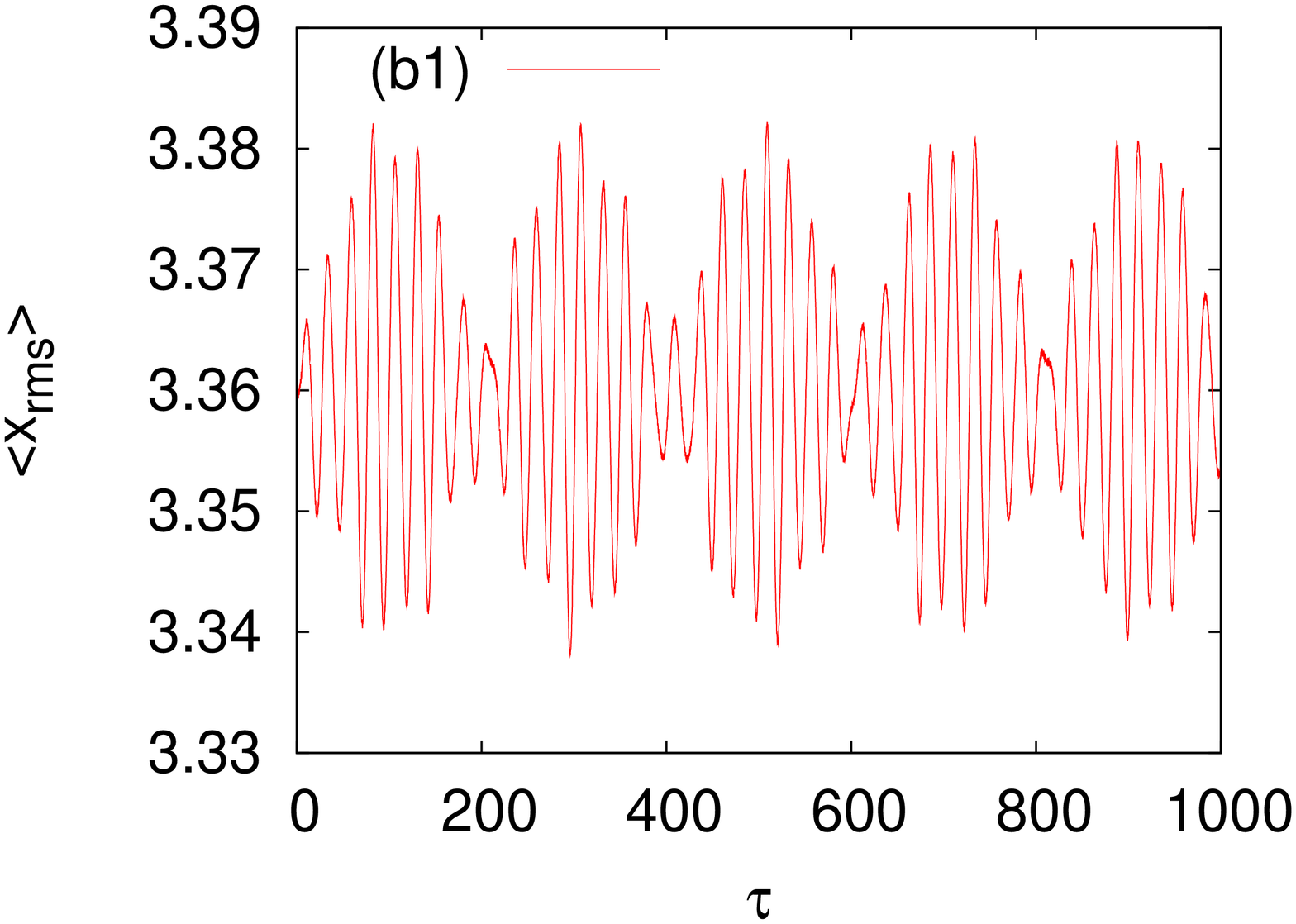}& \includegraphics [scale=0.3,angle=-90] {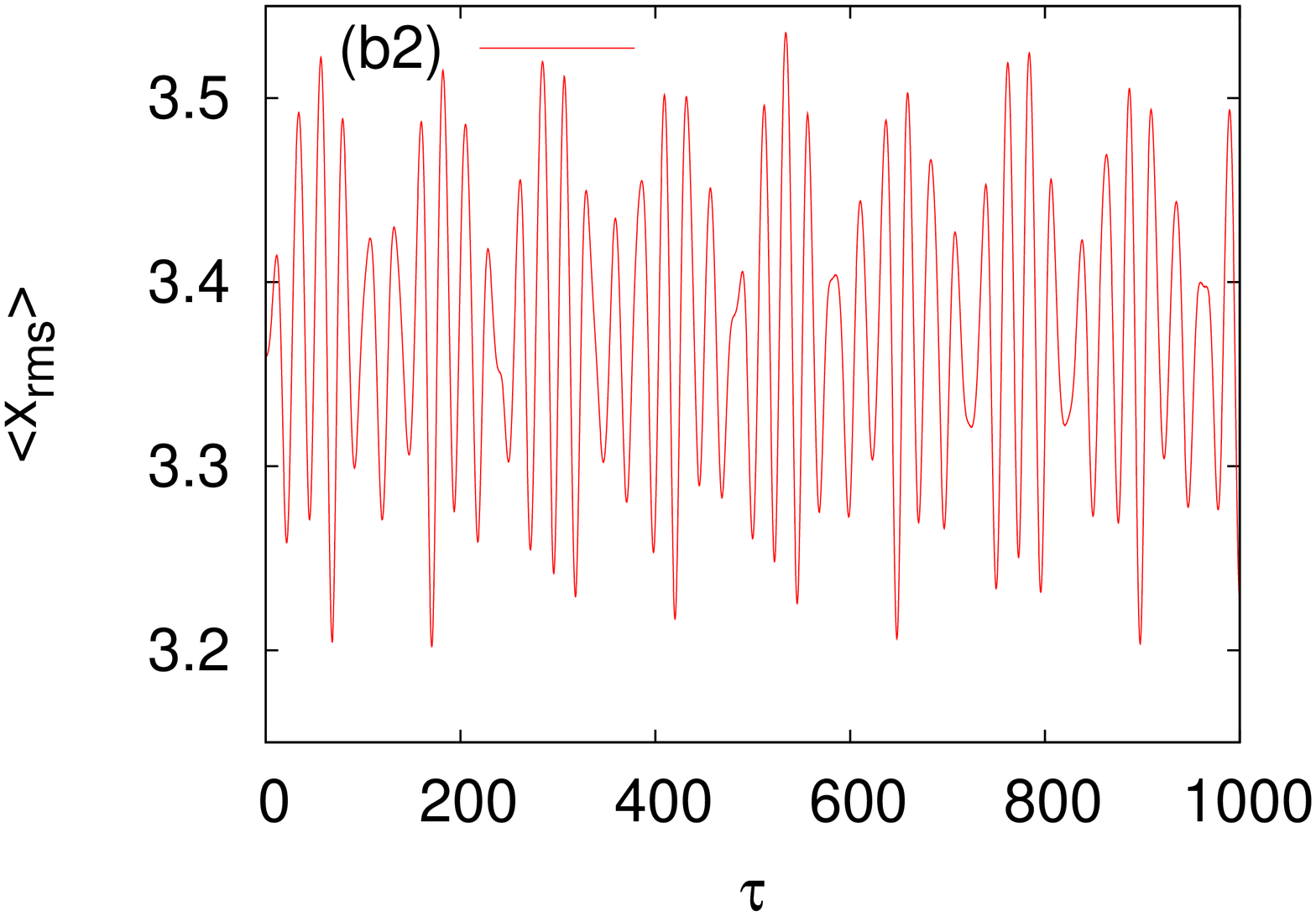}\\
\includegraphics [scale=0.3,angle=-90] {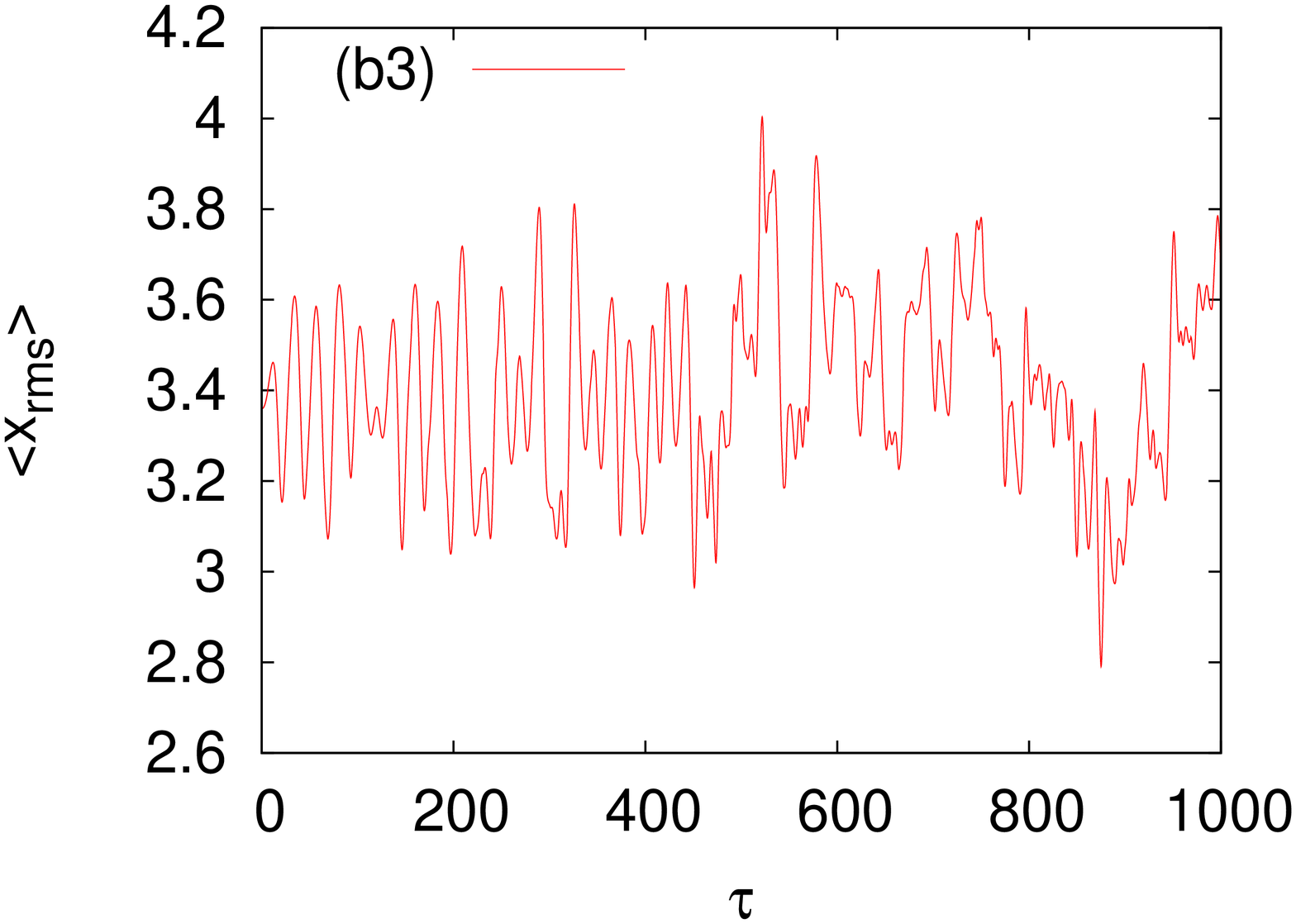} &\includegraphics [scale=0.3,angle=-90] {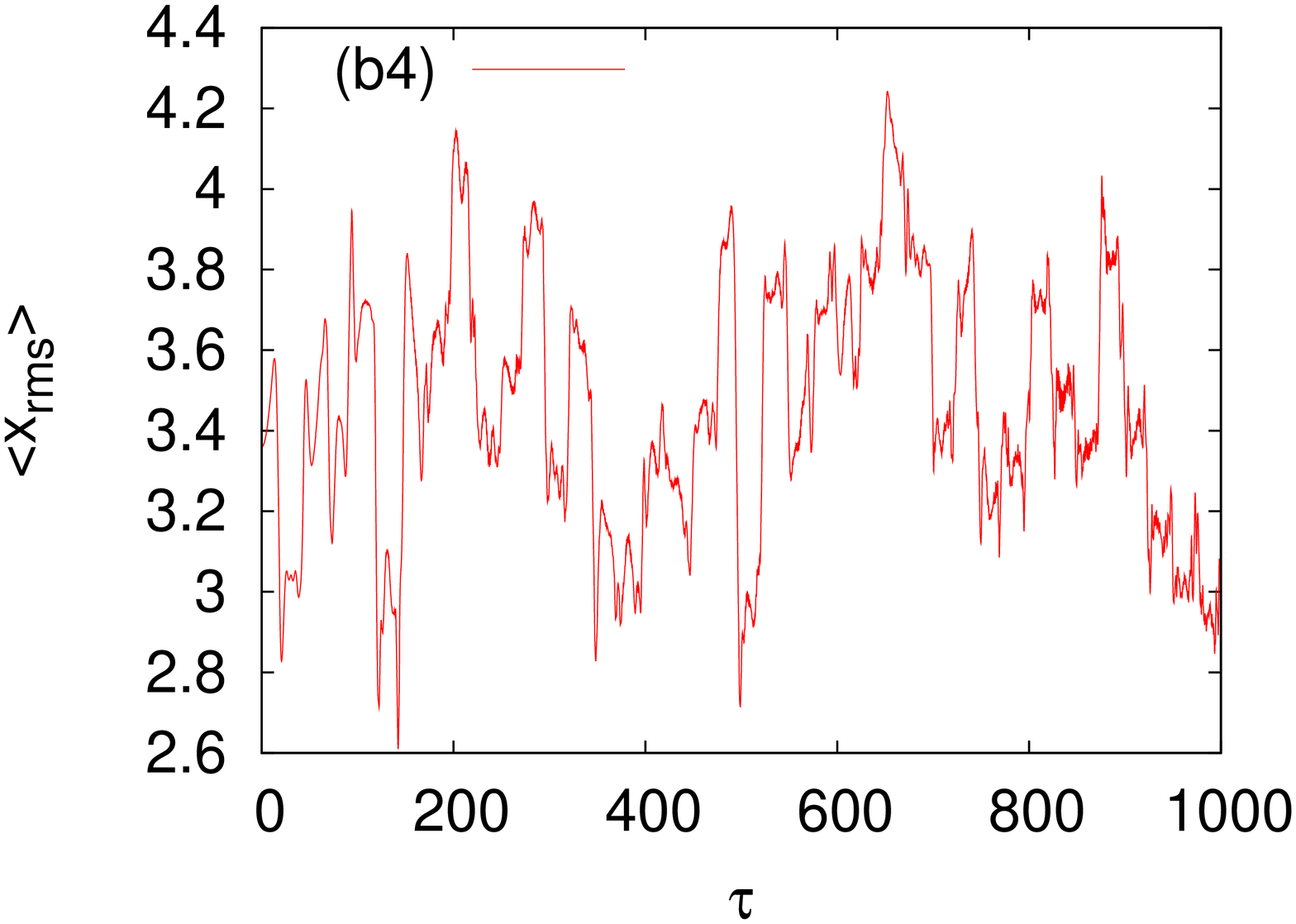}
\end{tabular}

\caption{Parametric resonances and instability in the width of the BEC vs dimensionless time $\tau$ in a combined harmonic trap and a time modulated optical lattice. The parameters are, $\gamma=0.1$, (b1): $\alpha=0.01$, (b2): $\alpha=0.1$,(b3): $\alpha=0.2$,(b4): $\alpha=0.5$. }

\label{figure2}
\end{figure}

\section{Numerical results}

In the numerical simulations, we have used the split-step Crank Nicolsan  method to solve the GP equation in one space variable. We have used the imagetime1D.F and realtime1D.F codes \citep{anand}. The ground state solution is found by imagetime1D.F code with $NPAS=100000$, $NRUN=20000$ with space step $dX=0.01$, time step $dt=0.0002$. To find the ground state solution, we have used the initial Gaussian wavefunction $\phi(X,\tau=0)=exp^{-X^{2}/2}/\sqrt{\pi}$ in the imagetine1D.F code. Then to study the time evolution of the GP equation, we have used the realtime1D.F code with $d \tau=0.001$. The results of the numerical simulations are now presented below.

To perform a systematic numerical study of the above Eqn. \ref{gp_4}, we take the value of $\overline{\lambda_{l}}=4.0$, $V_{o}=10.0$, $\omega_{r}=0.5$ . To perform the simulations we have first prepared the ground state wavefunction of BEC using the imaginary time propagation method with the $ V(X,\tau)$ at $\tau=0$. At $\tau=0$, $T=1$ and then this ground state is propagated in the real time with $V(X,\tau) =  \frac{\gamma X^2}{2} + AT sin^{2}\Big( \frac{2\pi X}{{\overline \lambda_{l}}}\Big) $. We have used the value of $dX$ $=$ $0.01$, $d\tau$ $=$ $0.0001$, $\eta_{1D}$ $=$ $17.0$. When the frequency of oscillation of the scattering length is an even multiple of the radial or axial natural oscillation  frequency, respectively, the radial or axial oscillation of the condensate exhibits resonance \citep{adhikari}.  In order to distinguish the present work from parametric resonances studied earlier \citep{adhikari}, we always consider the case when the frequency of oscillation of the optical lattice depth is not equal to the axial or radial natural oscillation frequency. The natural frequency of axial oscillations is $2 \omega$, so in order to stay away from resonance we have taken $\omega_{r}=0.5$ as mentioned above.

Fig.1 shows the time evolution of the $rms$ value of the condensate axial width $<X_{rms}>$ in a weak harmonic potential $\gamma=0.02$ for four different values of the perturbation strength $\alpha=0.01$ (plot a1), $\alpha=0.1$ (plot a2), $\alpha=0.2$(plot a3), $\alpha=0.5$ (plot a4).  We clearly see the competing effect of the harmonic potential and the periodic modulation of the intensity of the optical lattice. The harmonic potential tries to localize the condensate at the center of the trap while the periodic modulation of the lattice generates Bogoliubov excitations which with time grows and tends to delocalize the condensate. In plot $a1$, the periodically varying part of the lattice $\alpha \sin{\omega_{r} \tau}$ is weak and varies between zero and $\alpha$. The collapse and revival seen in plot $a1$ is a result of competition between the harmonic potential and $\alpha \sin{\omega_{r} \tau}$. Due to the weak value of the perturbing factor $\alpha$, the harmonic potential does not allow the condensate to become unstable and the balance between both the competing factors induces a continuous growth and decay cycles in the amplitude of $rms$ value of the condensate width. Now as we increase the perturbing strength $\alpha$ to $0.1$($a2$), parametric amplification of the Bogoliubov modes begin to appear after a certain time interval.  After a certain time, the harmonic potential is no longer able to stabilize and localize the condensate and there is an unprecedented growth of the parametric excitations and the condensate becomes unstable. On increasing $\alpha$ further to $0.2$ ($a3$) and $0.5$ (a4), the parametric amplification and hence instability of the condensate is more pronounced and the growth of the unstable modes starts almost immediately and the harmonic potential is completely unable to stabilize the condensate.

In Fig.2, we now show the dynamics of the $rms$ value of the condensate width in slightly stronger harmonic potential $\gamma=0.1$ for same values of $\alpha$, $\alpha=0.01$ (plot b1), $\alpha=0.1$ (plot b2), $\alpha=0.2$(plot b3), $\alpha=0.5$ (plot b4). Now since in Fig.2, the harmonic potential is stronger than that in Fig.1, naturally the value of $\alpha$ needed to make the condensate unstable and amplify the unstable modes is higher. At $\alpha=0.1$, we observe that the width still shows the typical growth and decay cycles as compared to the previous case of Fig.1, where at $\alpha=0.1$, the condensate becomes parametrically unstable after a certain time. In this case now, the condensate is completely unstable only when $\alpha \geq 0.2$. Another important point to note is that the $rms$ width of the condensate in the relatively stronger harmonic potential of Fig.2 is much smaller as compared to that in Fig.1.

The parametric instability observed in Fig.1 and Fig.2 for large value of $\alpha$ is a consequence of the exponential growth of the Bogoliubov modes of the system induced by the periodic variation of the optical lattice depth. In order for these Bogoliubov modes to be amplified, they must preexist at $t=0$. This means that the initial state is not a pure ground state. Indeed the system is certainly not in its ground state at $t=0$ due to excitations produced when loading the condensate into the lattice. Some numerical noise is always present and yields a very small occupation of Bogoliubov modes. The exponential growth of the parametric instability with time is due to the nonlinear mean-field dynamics of the condensate governed by the GP equation. This nonlinear dynamics is triggered by the growth of the parametrically unstable modes.

The growth of the $rms$ width with time due to the periodic variation of the optical lattice also motivated us to analyze how this periodic modulation influences the spatial extension of the condensate density. Fig.3 and Fig.4 show the density plots corresponding to Fig.1 and Fig.2 respectively. Clearly as the strength of the perturbation $\alpha$ increases, the wavefunction also gets progressively delocalized and starts extending over more lattice sites. On increasing the strength of the harmonic potential (Fig.4), the influence of $\alpha$ in delocalizing the condensate wavefunction is less effective, since a stronger harmonic potential now localizes the condensate to a greater extent and acts against the perturbation. The periodic modulation redistributes the atoms in the lattice wells in a random manner and hence the nonuniform density distribution across the lattice wells for large $\alpha$ is seen in Fig.3 and Fig.4.

\begin{figure}[t]

\begin{tabular}{cc}
\includegraphics [scale=0.3,angle=-90] {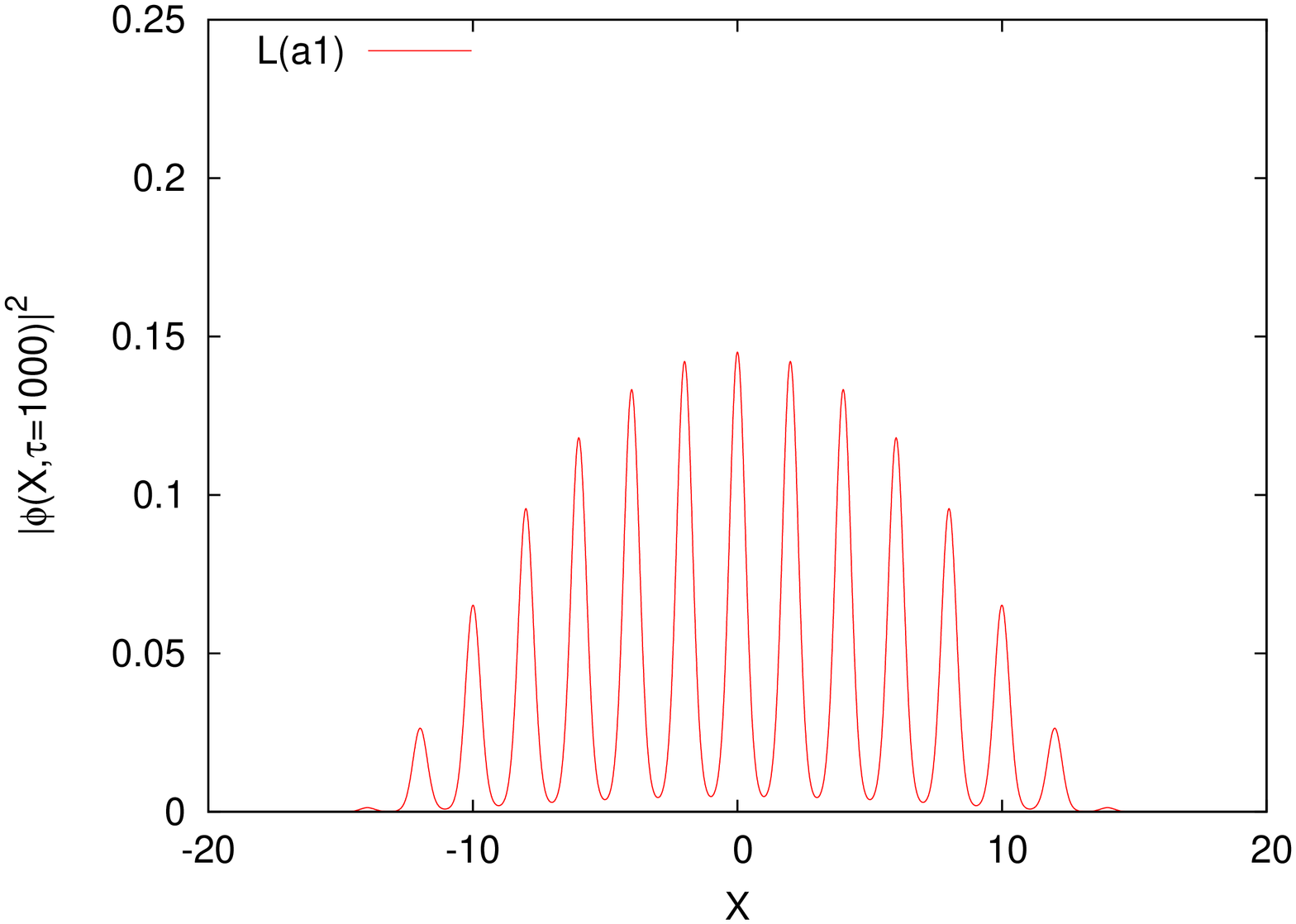}& \includegraphics [scale=0.3,angle=-90] {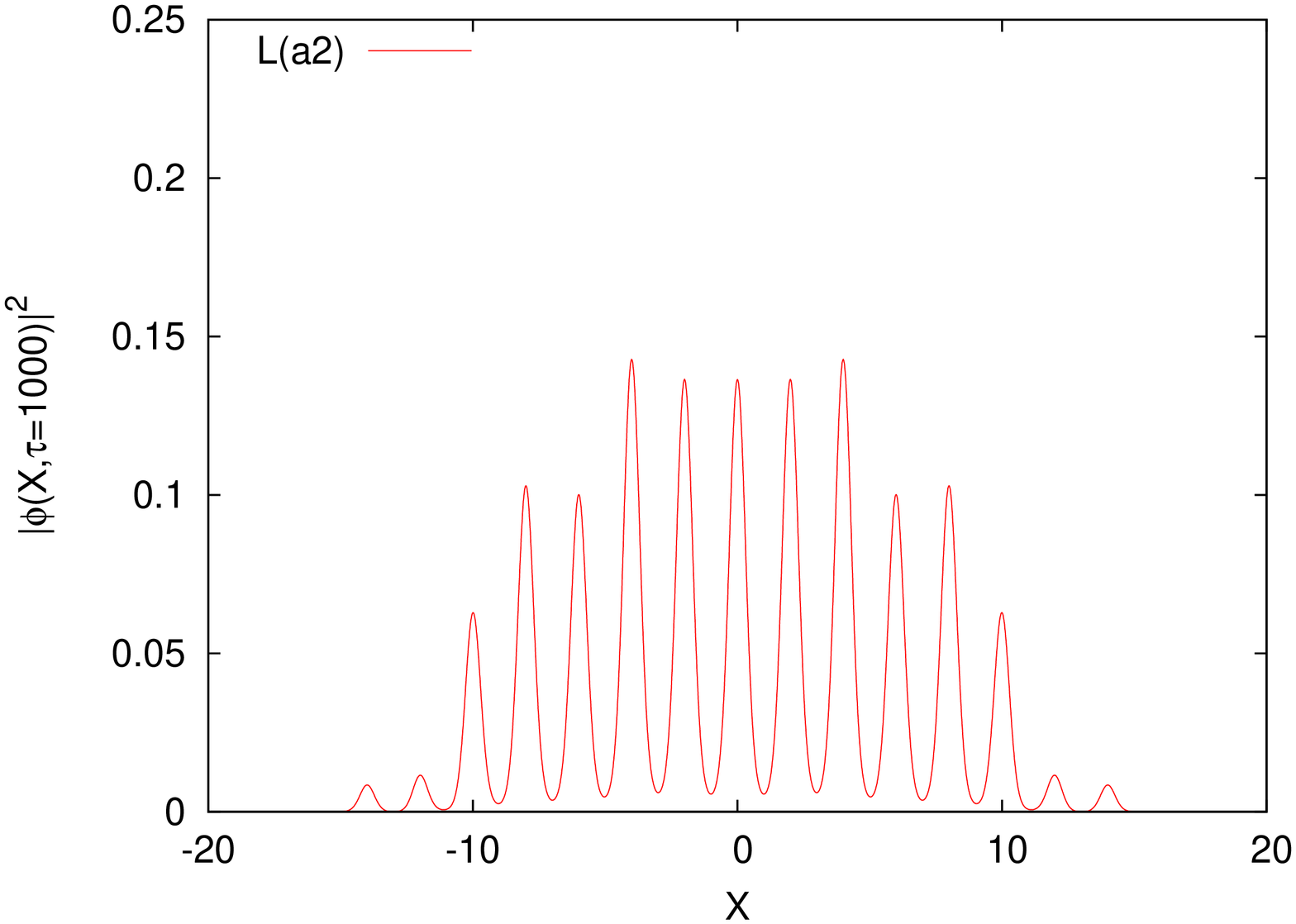}\\
\includegraphics [scale=0.3,angle=-90] {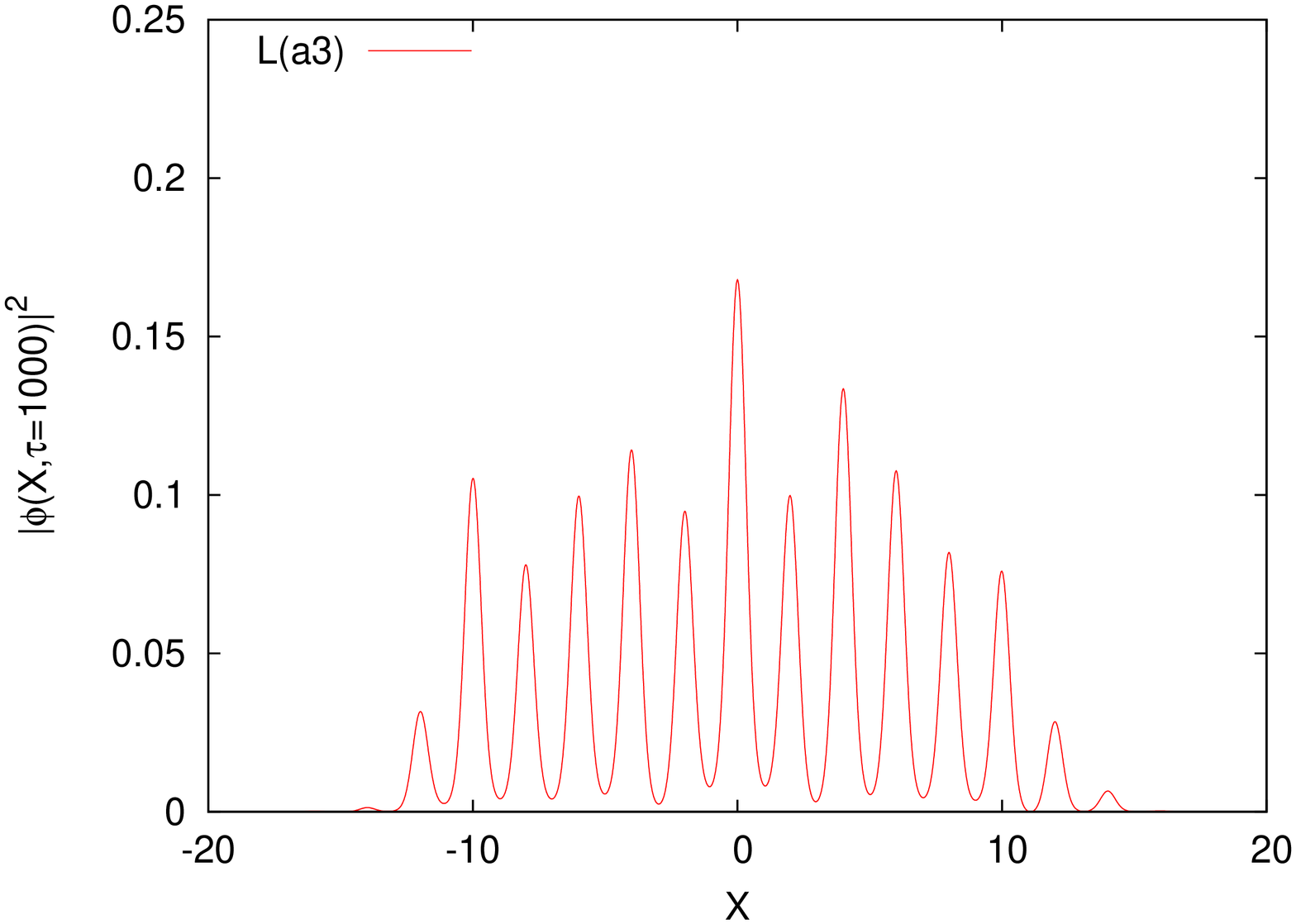} &\includegraphics [scale=0.3,angle=-90] {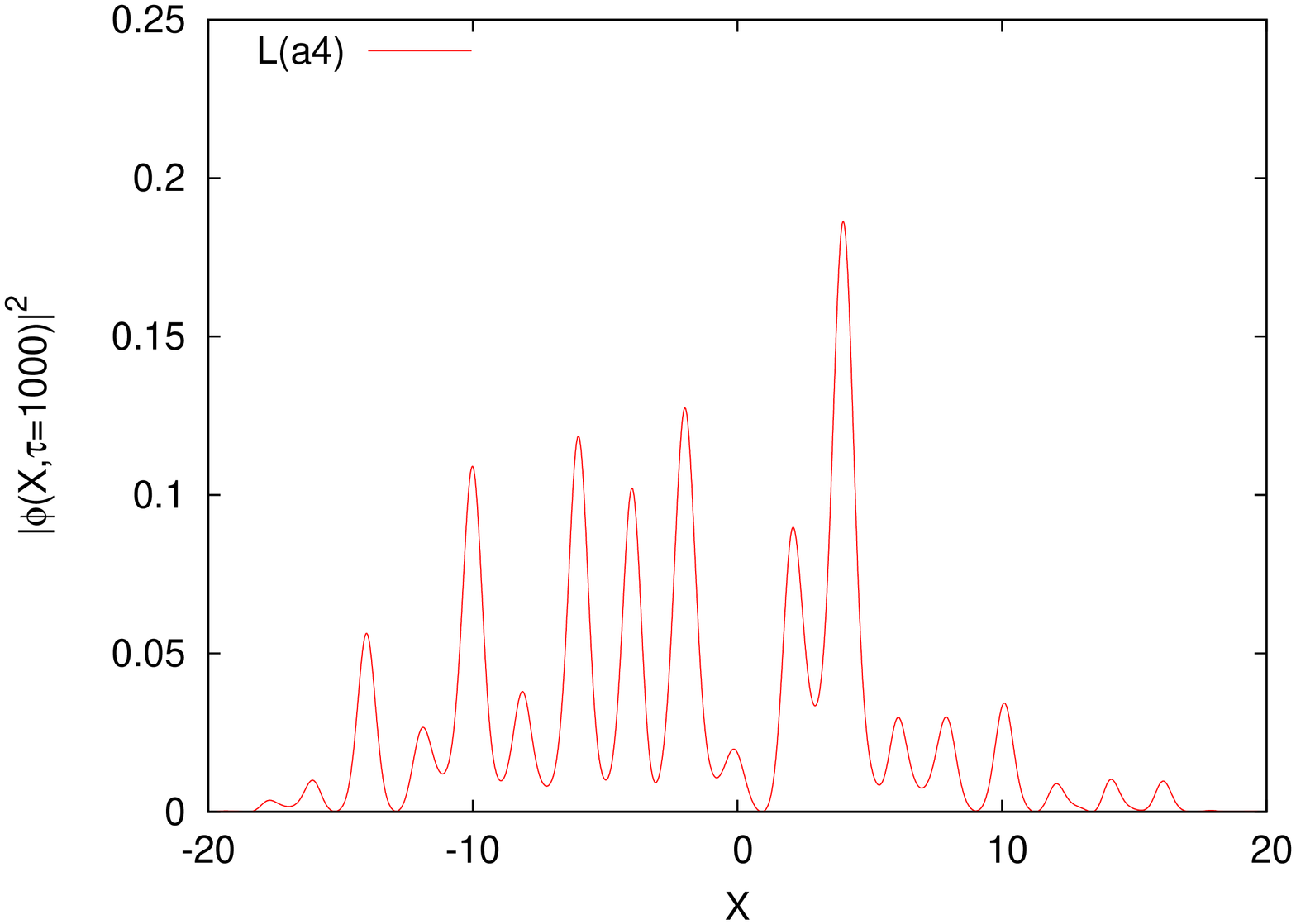}
\end{tabular}

\caption{Density distribution $|\phi(X)|^{2}$ at $\tau=1000$ vs $X$ corresponding to the cases considered in Fig.1. The parameters are $\gamma=0.001$, $L(a1)$: $\alpha=0.01$, $L(a2)$: $\alpha=0.1$, $L(a3)$: $\alpha=0.2$, $L(a4)$: $\alpha=0.5$. }

\label{figure3}
\end{figure}

\begin{figure}[t]

\begin{tabular}{cc}
\includegraphics [scale=0.3,angle=-90] {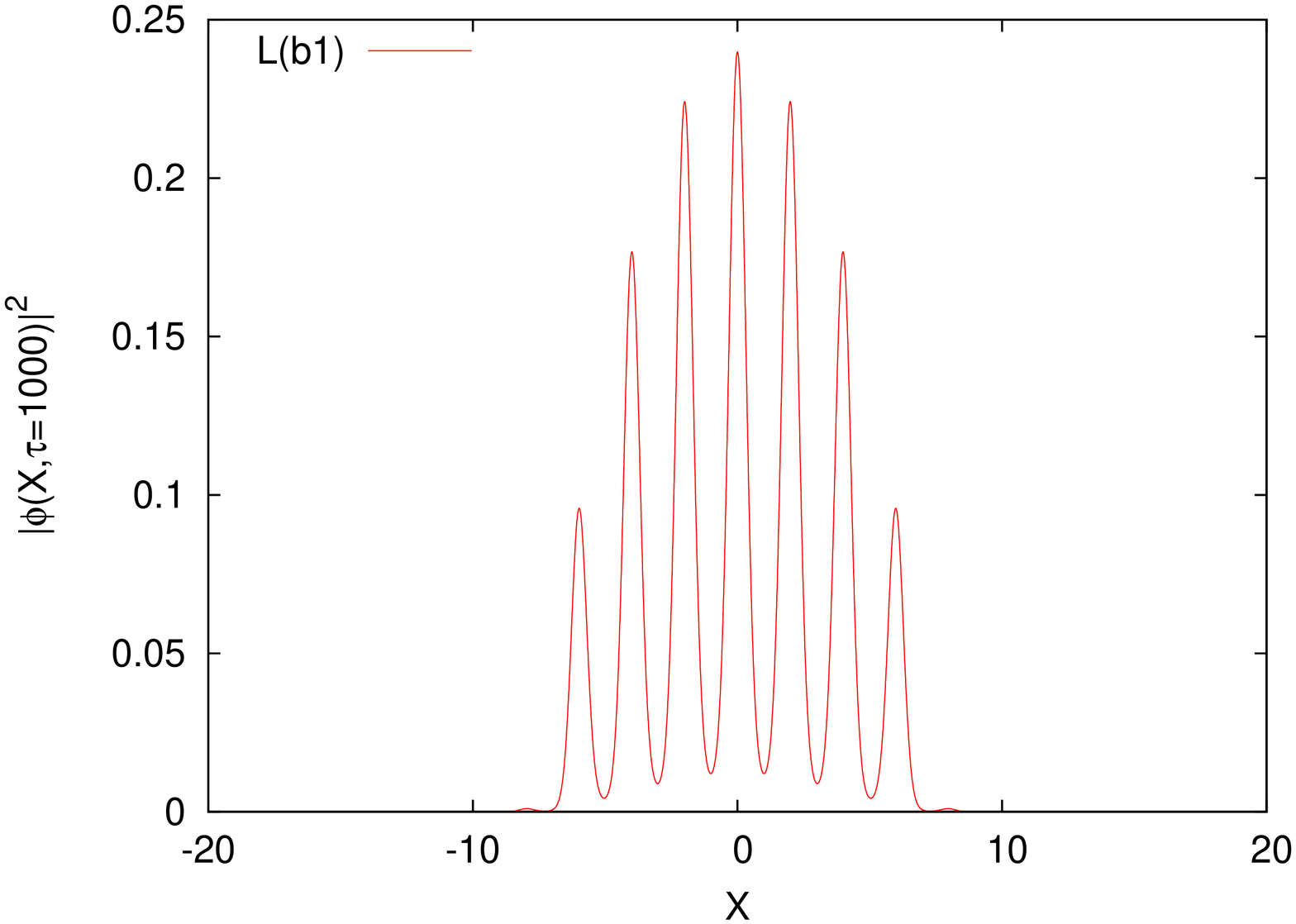}& \includegraphics [scale=0.3,angle=-90] {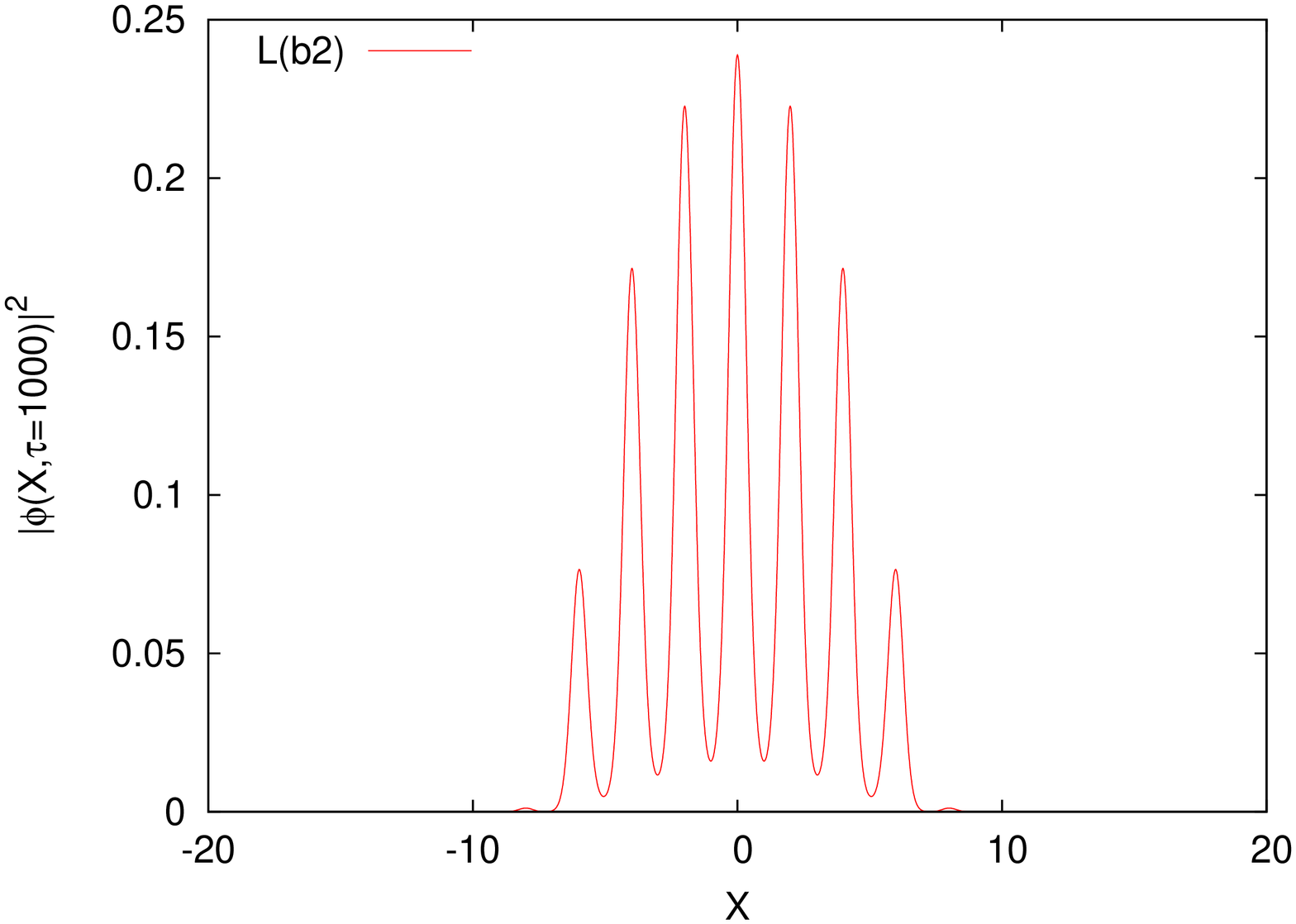}\\
\includegraphics [scale=0.3,angle=-90] {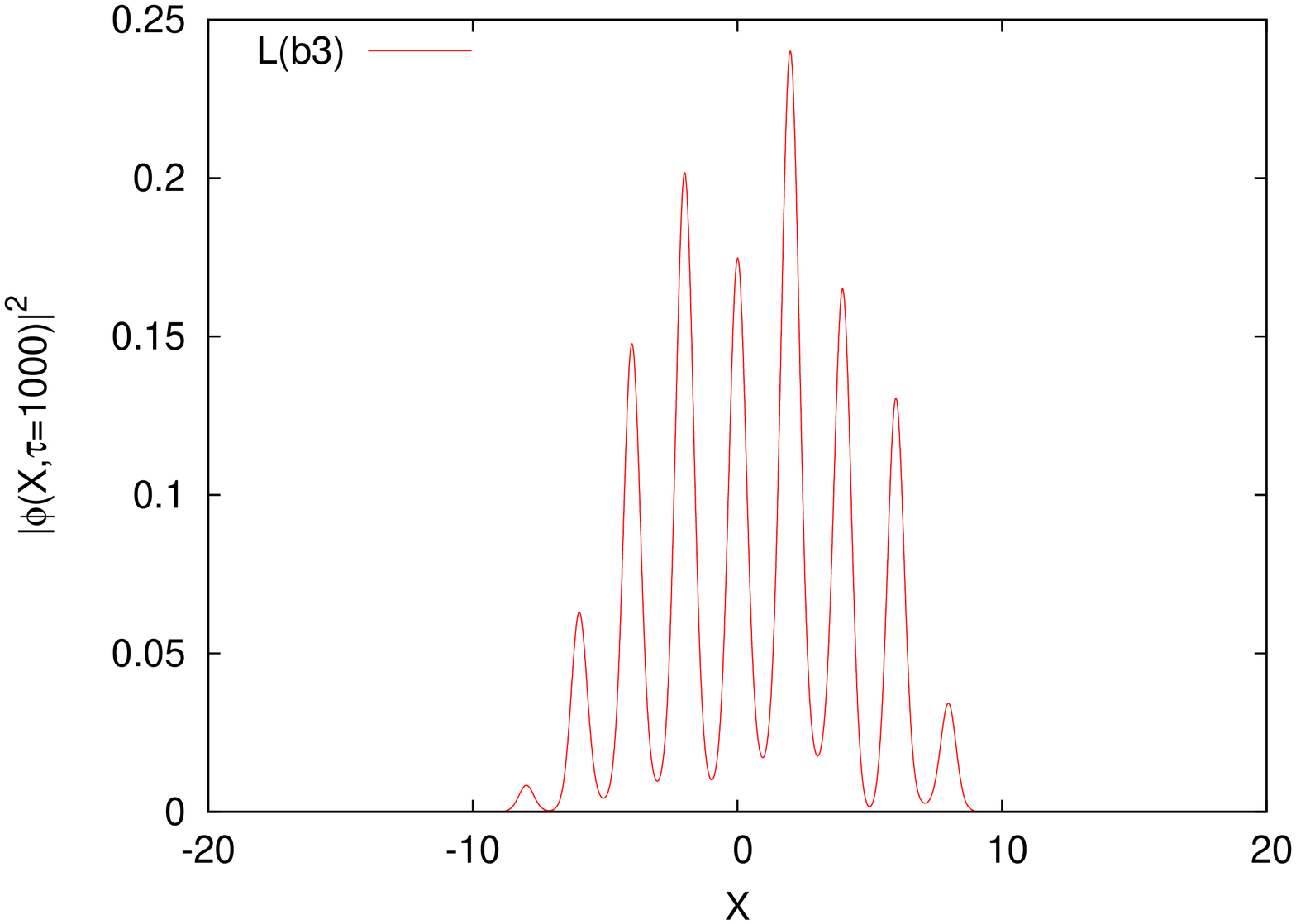} &\includegraphics [scale=0.3,angle=-90] {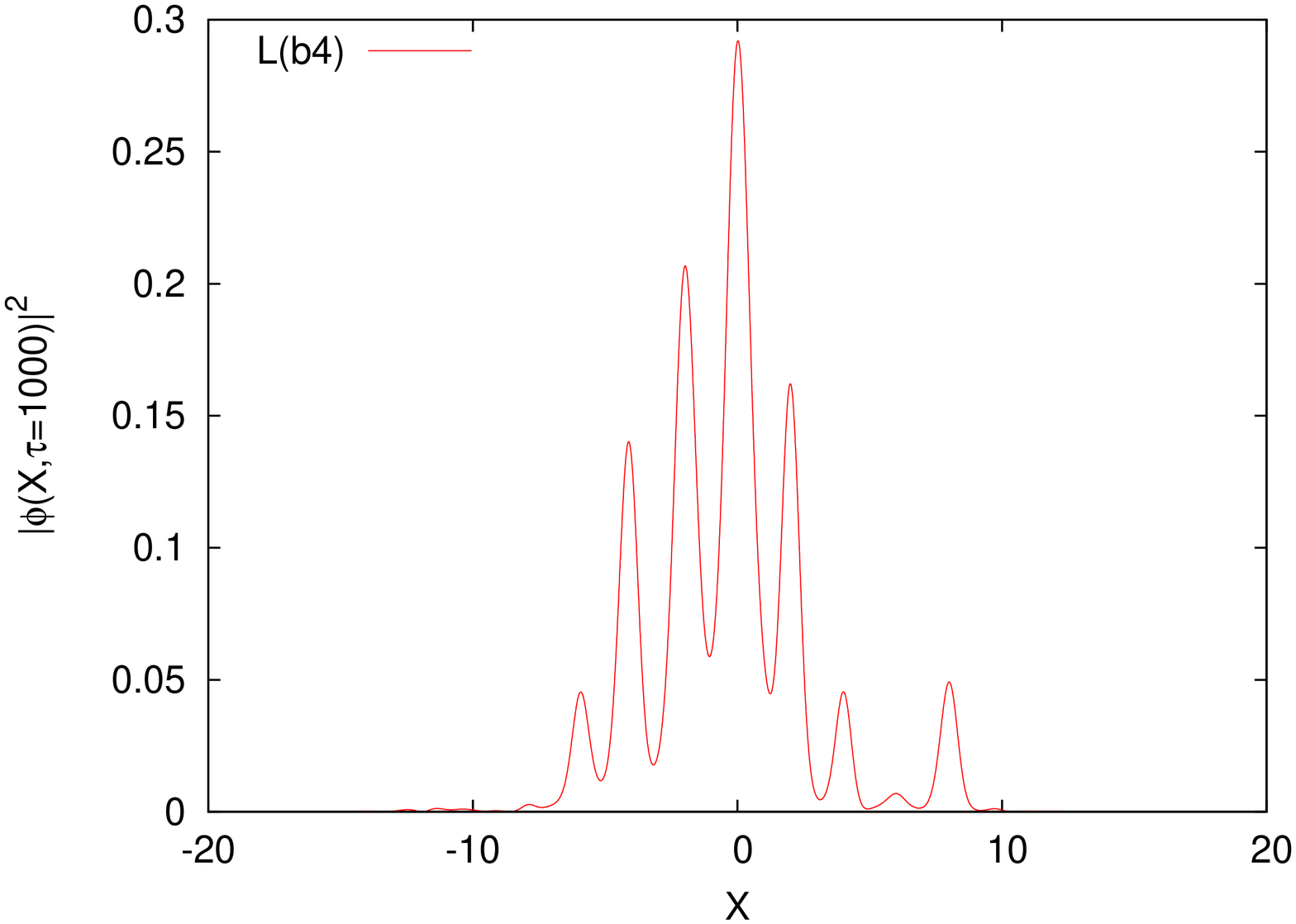}
\end{tabular}

\caption{Density distribution $|\phi(X)|^{2}$ at $\tau=1000$ vs $X$ corresponding to the cases considered in Fig.1. The parameters are $\gamma=0.1$, $L(b1)$: $\alpha=0.01$, $L(b2)$: $\alpha=0.1$, $L(b3)$: $\alpha=0.2$, $L(b4)$: $\alpha=0.5$.  }

\label{figure4}
\end{figure}

\section{Variational Analysis}

\begin{figure}[t]

\begin{tabular}{cc}
\includegraphics [scale=0.86] {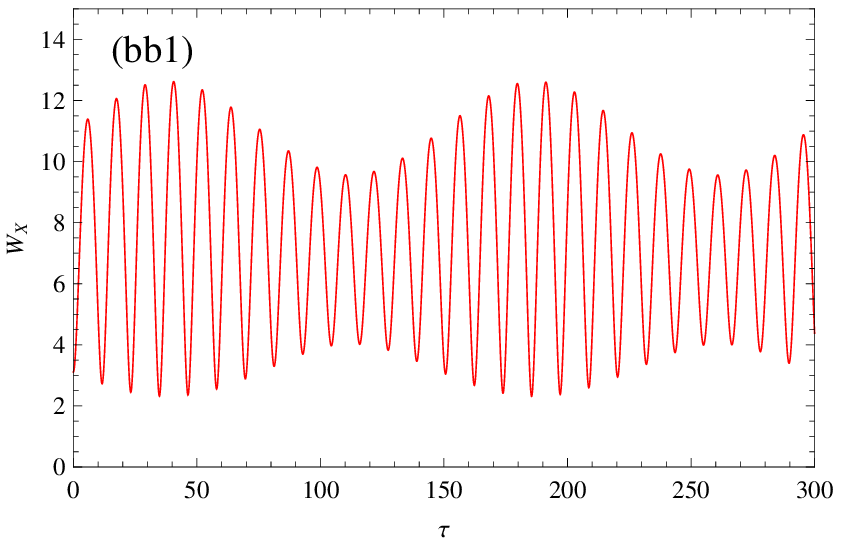}&\includegraphics [scale=0.86] {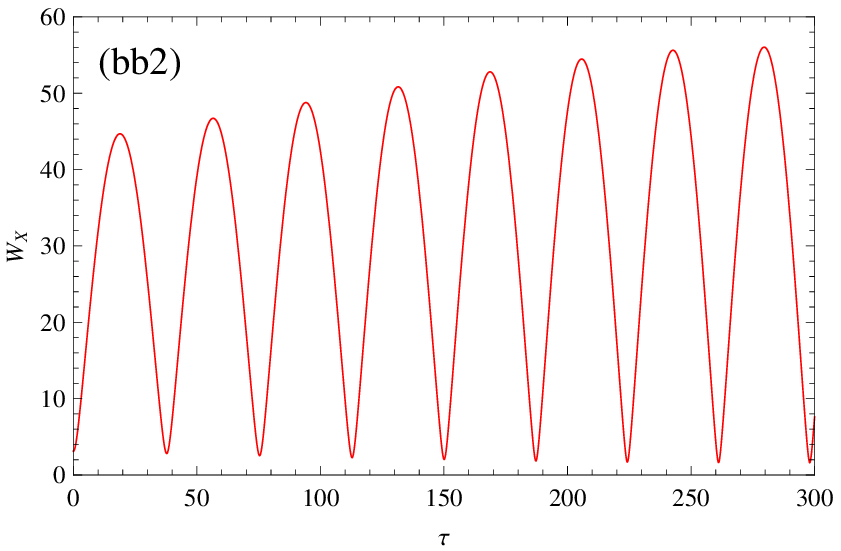}\\
\includegraphics [scale=0.85] {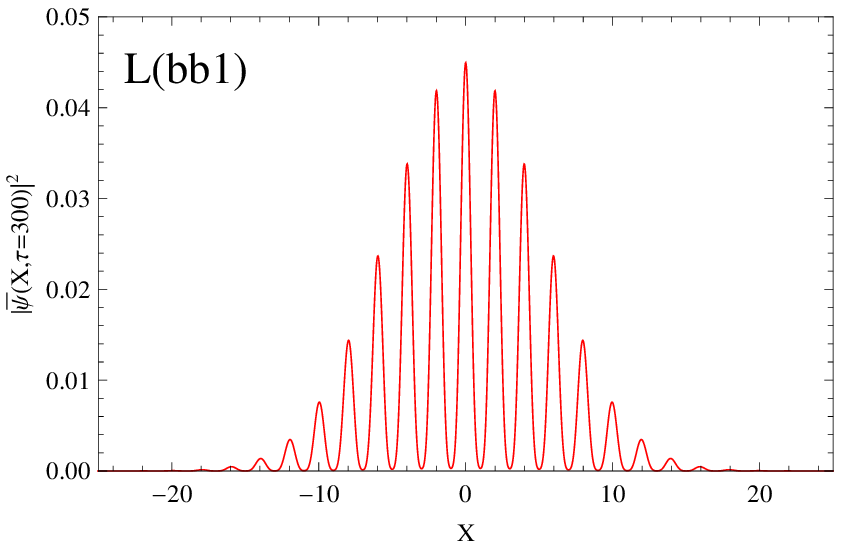}&\includegraphics [scale=0.9] {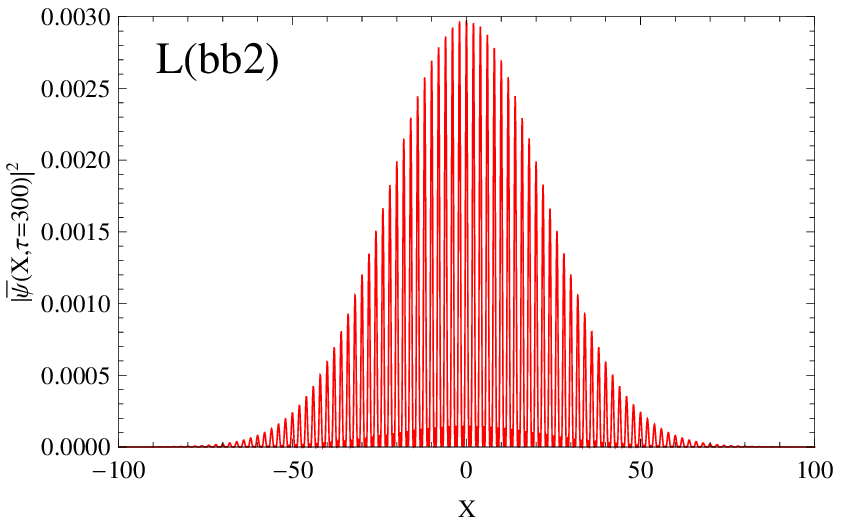}
\end{tabular}

\caption{Analytical results of the BEC width (plots bb1 and bb2) vs dimensionless time $\tau$ and density $|\overline \psi(X,\tau)|^{2}$ (L(bb1) and L(bb2) at $\tau=300$ vs $X$. The parameters are $\alpha=0.2$, $\gamma=0.1$ (plots bb1 and L(bb1)) and $\gamma=0.008$ (plots bb2 and L(bb2)). Here $\overline \psi(X,\tau)=\frac{\sqrt{\pi}\psi{X,\tau}}{N}$.}

\label{figure5}
\end{figure}

All the results that we have presented in the previous section are fully numerical. In this section, using the Lagrange variational technique, we reproduce the essential physics presented in the previous section.

The dynamics of a trapped one-dimensional repulsive Bose gas is described in the framework of the 1D Gross-Pitaeviskii equation

\begin{eqnarray}\label{gp_var1}
i\ \hbar\frac{\partial\psi(x,t)}{\partial t} &=& \Big[\frac{-\hbar^{2}}{2m}\frac{\partial^2}{\partial x^{2}} + V(x,t) + \eta_{1D}\vert\psi(x,t)\vert^{2}\Big] \psi(x,t)
\end{eqnarray}

with the total number of atoms $N=\int |\psi|^{2} dx$. Eq. \ref{gp_var1} is obtained for a highly elongated cigar shaped BEC. In order to create a cigar shaped elongated BEC, the frequency of the harmonic trap along the transverse direction should be made larger than the one in the axial (along the direction of the optical lattice) direction. The condition for 1D approximation is $\omega_{y},\omega_{y}$ $\gg$ $\omega_{x}$, $\mu$ $\ll$ $\hbar \omega_{y}, \hbar \omega_{z}$, where $\mu$ is the chemical potential. To analytically solve  Eq. \ref{gp_var1}, we need a trial wavefunction. A proper choice of the trial wavefunction is very important in this technique, so we are assuming that the total wavefunction is a
gaussian envelope multiplied by a periodic part. The periodic part of the wavefunction which arises from the optical lattice consists of a unmodulated and a modulated part.  The modulated part of the periodic wavefunction is small compared to the unmodulated part because the optical lattice modulates the  gaussian wavepacket along the longitudinal direction only slightly. To this end, we use the following trial wavefunction for the periodic part of the wavefunction

\begin{eqnarray}\label{waveeqn_2}
\vert \psi\vert =  \Big[1 + \sigma_{2}\   T' cos\Big(\frac{4\pi x}{\lambda_{l}}\Big) \Big]^{\frac{1}{2}},
\end{eqnarray}

where $A' = V$ and $T'=1+\alpha \sin(\Omega t)$ and $\sigma_{2}$ is a variational parameter.

Using the above ansatz Eq. \ref{waveeqn_2}, we can write the trial wavefunction as

\begin{eqnarray}\label{waveeqn_3}
\psi(x,t) =\frac{\ \ \ N e^{i \phi}}{\sqrt{\pi}\  W_{x}}\     \  \Big[1 + \sigma_{2}\   T' cos\Big(\frac{4\pi x}{\lambda_{l}}\Big) \Big] \        exp\Big[-\frac{(x-x_{0}^{2})}{2 W^{2}_{x}} + ix\alpha_{x} + ix^{2} \beta_{x}\Big],
\end{eqnarray}

where $W_{x}$ is the width of the condensate, $\alpha_{x}$ is the slope, $\beta_{x}$  is the (square root of the curvature radius)$^{-1}$, $x_{0}$ is the center of the cloud and $\phi$ is the global phase are the variational parameters.  Above trial wavefunction can be normalized with respect to $N$ in the limit $e^{-\frac{jW_{x}^{2}}{\overline \lambda^{2}}}$ $\rightarrow$ $0$, where $j$ is any positive integer.  We are taking this approximation in all the calculations because $\overline \lambda$ is very small as compared to $W_{x}$. This leads us to the final trial wavefunction

\begin{eqnarray}
\psi(x,t) =\frac{\ \ \ N e^{i \phi}}{\sqrt{\pi}\  W_{x}}\  \Big[1 + \frac{(\sigma_{2} T')^{2}}{\ \ 2}    \Big]^{\frac{-1}{2}}\ \Big[1 + \sigma_{2}\   T' cos\Big(\frac{4\pi x}{\lambda_{l}}\Big) \Big] \        exp\Big[-\frac{(x-x_{0}^{2})}{2 W^{2}_{x}} + ix\alpha_{x} + ix^{2} \beta_{x}\Big]
\end{eqnarray}

The corresponding 1D Lagrangian density is

\begin{eqnarray}
\pounds(t) = \frac{i\  \hbar }{2}(\psi \frac{\partial\psi^{*}}{\partial t} - \psi^{*} \frac{\partial\psi}{\partial t})
\ +\ \frac{\  \hbar^2}{2m} \vert \frac{\partial \psi}{\partial x}  \vert^2  +  V(x,t) \vert \psi \vert^{2} + \eta_{1D} \vert \psi \vert^{4}
\end{eqnarray}

 The 1D Lagrangian of the system is found from as

\begin{eqnarray}
L_{1D} = <\pounds>=\int^{\infty}_{-\infty} \pounds \ dx.
\end{eqnarray}

This yields,

\begin{eqnarray}
L_{1D} (t) &=&  \ \Big( \frac{W_{x}^{2}}{2}+x_{o}^{2} \Big) \Big( \hbar N \dot{\beta_{x}} + \frac{2 \hbar^{2}}{m}N \beta^{2}_{x} + \frac{\gamma}{2}m N \omega_{x}^{2}  \Big)  +  \hbar N \dot{\phi}+
x_{o} \Big( \hbar N \dot{\alpha_{x}} + \frac{2\hbar^{2}N}{m}\alpha_{x}\beta_{x}  \Big) \nonumber \\ &+&  \frac{\hbar^{2} N}{2m} \Big( \frac{1}{2W_{x}^{2}} + \alpha_{x}^{2} \Big) + \frac{\eta_{1D} N^{2} f_{1}(A' T')}{\sqrt{2\pi} W_{x}} +\ f_{2}(A' T'),
\end{eqnarray}

where $A'$ ,$T'$ are the same as already described and

\begin{eqnarray}
f_{1}(A'T')=\Big[1+\frac{(\sigma_{2}T' )^{2}}{2} \Big]^{-2}\   \Big[ 1 + 3(\sigma_{2} T')^{2} + \frac{3}{8} (\sigma_{2} T')^{4}   \Big]
\end{eqnarray}

\begin{eqnarray}
f_{2}(A'T')=N\Big[\frac{A'T'}{2} -\frac{A'T'}{2}\sigma_{2}T'\Big(1+\frac{(\sigma_{2}T')^{2}}{2}\Big)^{-1}\Big] + \frac{\hbar^{2}N}{2m} \Big[\frac{1}{4} \Big(\frac{4\pi}{\lambda_{l}}\Big)^{2} (\sigma_{2}T')^{2} \Big( 1+\frac{(\sigma_{2}T')^{2}}{2} \Big)^{-1}  \Big]
\end{eqnarray}

The evolution of the Lagrange parameters, $q_{i}\equiv(W_{x} , x_{o} , \alpha_{x} ,\beta_{x},\sigma_{2}  )$ is ruled by the corresponding set of Lagrange equations

\begin{eqnarray}
\frac{d}{dt}\Big[ \frac{\partial L_{1D}}{\partial \dot{q_{i}}} \Big]\ =\ \frac{\partial L_{1D}}{\partial q_{i}},
\end{eqnarray}

which give us equation of motion for the conservation of the norm,

\begin{equation}
\frac{dN}{dt}=0,
\end{equation}

the equation of motion for the slope gives,
\begin{eqnarray}
\hbar \dot{x_{0}} = \frac{\hbar^{2} \alpha_{x}}{m} + \frac{2\hbar^{2}}{m} \beta_{x} x_{0},
\end{eqnarray}

the equation of motion for the (curvature radius)$^{-\frac{1}{2}}$ gives,

\begin{eqnarray}
\hbar N (W_{x} \dot{W_{x}}   +    2 x_{0} \dot{x_{0}} ) = \frac{4 \hbar^{2} N}{m} \beta_{x} ( \frac{W_{x}^2}{2} + x_{0}^{2} )  +  \frac{2 \hbar^{2} N }{m} \alpha_{x} x_{0},
\end{eqnarray}

the equation of motion for the center of cloud gives,

\begin{eqnarray}
2 x_{0} ( \hbar N \dot{\beta_{x}}  + \frac{2 \hbar^{2} N}{m}  \beta_{x}^{2}   + \frac{\gamma}{2} N m \omega_{x}^{2} ) + ( \hbar N \dot{\alpha_{x}} +\frac{2 \hbar^{2} N}{m} \alpha_{x} \beta_{x})  = 0,
\end{eqnarray}

and for width gives,

\begin{eqnarray}
W_{x}(\hbar N \dot{\beta_{x}} + \frac{2 \hbar^{2} N}{m}  \beta_{x}^{2} +\frac{\gamma}{2} N m \omega_{x}^{2} )) = \frac{\hbar^{2} N }{2mW_{x}^{3}} +  \frac{\eta_{1D} N^{2}}{\sqrt{2 \pi} W_{x}^{2}} f_{1}(A'T').
\end{eqnarray}

The equation of motion for the parameter $\sigma_{2}$ yields,
\begin{eqnarray}
\sigma_{2}=\frac{A'/2}{\Big[   \frac{4\eta_{1D}N}{\sqrt{2\pi}W_{x}} + \Big(\frac{4\pi}{\lambda_{l}}\Big)^{2} \frac{\hbar^{2}}{4m} \Big]}
\end{eqnarray}

In calculating the above equation of motion we have assumed that $\sigma_{2}<<1$ so that we have ignored all higher order terms.

On rearranging the above equations we get evolution of the slope and curvature as

\begin{eqnarray}
\beta_{x}=\frac{m}{2\hbar} \frac{\dot{W_{x}}}{W_{x}}
\end{eqnarray}

\begin{eqnarray}
\alpha_{x}=\frac{m}{\hbar} \dot{x_{o}} - 2\beta_{x} x_{o}
\end{eqnarray}

and finally the evolution of width as

\begin{eqnarray}
\ddot{W_{x}}+\gamma \omega^{2}_{x} W_{x} \ =\ \frac{\hbar^{2}}{m^{2}} \frac{1}{W_{x}^{3}} + \frac{2 \eta_{1D}N f_{1}(A' T')}{\sqrt{2\pi}mW_{x}^{2}}
\end{eqnarray}

where $\eta_{1D} = \frac{2 \hbar^{2} a}{m a_{\perp}^{2}}$.  Let us introduce the constant $ P=\sqrt{\frac{2}{\pi}}N\frac{a}{l}$ where P denotes
the strength of interaction between two atoms. After making the transformations $ t\rightarrow \frac{\tau}{\omega} \ , \ x\rightarrow Xl \ ,\ y\rightarrow Yl \ , \ z\rightarrow Zl \ ,\ l=\sqrt{\frac{\hbar}{m\omega}}$, the dimensionless equation of motion for the width $W_{X}$ leads to

\begin{eqnarray}\label{width_1}
\ddot{W_{X}}+\gamma W_{X}=\frac{1}{W_{X}^{3}}+\Big[ 2P f_{1}(AT) (\frac{l}{a_{\bot}} )^{2}   \Big] \frac{1}{W_{X}^{2}}
\end{eqnarray}

where $a_{\bot}=\sqrt{   \frac{\hbar}{m\omega_{\bot}}   }$ in the limit $\omega_{\bot}>>\omega_{x}$, $\omega_{\bot}=\omega_{y},\omega_{z}$. Note that $\omega$ $=$ $\omega_{x}$ since $\nu=1$.

  While solving the differential equation of width we have taken $P=0.07$, $N=460$, $\eta_{1D}=17.0$ where $a=5.1nm$ for the rubidium atoms , $\omega_{\bot}=100\omega_{x}$. In Fig.6, we show the time dynamics of the width calculated from Eqn. \ref{width_1} in a harmonic potential of strength $\alpha=0.2$ for two values of $\gamma=0.1$ (left plot) and $\gamma=0.008$ (right plot). Qualitatively we see the similarity with our numerical results. For strong value of $\gamma=0.1$, we clearly see the collapse and revival in the oscillations of the width while for a weaker value of $\gamma=0.008$, the parametric instabilities start to grow and the width of the condensate increases very fast with time. The density plots corresponding to $bb1$ and $bb2$ are shown in $L(bb1)$ and $L(bb2)$ respectively. Clearly we see that the wavefunction for the case $\gamma=0.008$ is more delocalized as compared to the case for $\gamma=0.1$. This result is again in agreement with numerical results. The occurrence of this type of parametric instability associated with Bogoliubov excitation in Bose-Einstein condensates originates from their superfluid nature and the existence of undamped collective excitations.

\section{Conclusion}

In this work we have explored the parametric excitations in an elongated Bose-Einstein condensate subject to a combined harmonic potential and $1D$ optical lattice whose intensity is periodically modulated in time. On one hand the periodic modulation generates Bogoliubov excitations which grows with time and tends to increase the condensate size and hence delocalize it and on the other hand the harmonic trap tries to localize the condensate at the center of the trap and act against the instability of the Bogoliubov modes. The ensuing nonlinear dynamics leads to parametric resonances (collapse and revival of the oscillations of the condensate width). The parametric resonances and instability are nicely accounted for by the numerical and analytical calculations.

\section{Acknowledgements}

One of the authors Priyanka Verma thanks the University Grants Commission, New Delhi for the Junior Research Fellowship. The authors thank P. Murugandam for discussions on the numerical simulations.


\begin{thebibliography}{plain}

\bibitem{castin}
Y. Castin and R. Dumm, Phys. Rev. Letts., {\textbf{79}}, 3553, (1997).
%
\bibitem{kagan}
Yu. Kagan, L. A. Maksimov, Phys. Rev. A, {\textbf{64}}, 053610, (2001).
%
\bibitem{kevrekidis}
P. G. Kevrekidis, A. R. Bishop and K. O. Rasmussen, J. Low Temp. Phys.,{\textbf{120}}, 205, (2000).
%
\bibitem{Engels07}
P. Engels, C. Antherton and M.A. Hoefer, Phys. Rev. Letts., {\textbf{98}}, 095301, (2007).
%
\bibitem{adhikari}
S.K Adhikari, J. Phys. B, {\textbf{36}}, 1109, (2003).
%
\bibitem{abdul1}
F.K. Abdullaev, R. M. Galimzyanov and K. Ismatullaev, Phys. Letts., {\textbf{357}}, 48, (2006).
%
\bibitem{abdul2}
F.K. Abdullaev, R. M. Galimzyanov and K. Ismatullaev, J. Phys. B, {\textbf{41}}, 015301, (2008).
%
\bibitem{stalinus02}
K. Staliunas et. al., Phys. Rev. Letts., {\textbf{89}}, 210406 , (2002).
%
\bibitem{stalinus04}
K. Staliunas et. al., Phys. Rev. A., {\textbf{70}}, 011601(R), (2004).
%
\bibitem{nicol}
A. I. Nicolin, R. Carretero-Gonz\'{a}lez and P G. Kevrekidis, Phys. Rev. A., {\textbf{76}}, 063609, (2007)
%
\bibitem{bhattacherjee}
A. Bhattacherjee, Phys. Scr. {\textbf{78}},045009,(2008).
%
\bibitem{tozzo1}
C. Tozzo, M. Kr\"{a}mer and F. Dalfovo, Phys. Rev. A., {\textbf{72}}, 023613, (2005).
%
\bibitem{tozzo2}
M. Kr\"{a}mer, C. Tozzo and F. Dalfovo, Phys. Rev. A., {\textbf{71}}, 061602(R), (2005).
%
\bibitem{stoferle}
T. St\"{o}ferle at al., Phys. Rev. Lett. {\textbf{92}}, 130403, (2004).
%
\bibitem{schori}
C. Schori et al., Phys. Rev. Lett. {\textbf{93}}, 240402, (2004).
%
\bibitem{victor}
V. M. P\'{e}rez-Garcia et al., Phys. Rev. A., {\textbf{56}}, 1424, (1997).
%
\bibitem{juan}
J. J. G. Ripoll and V. M. P\'{e}rez-Garcia et al., Phys. Rev. A., {\textbf{59}}, 2220, (1999).
%
\bibitem{mark}
M. Edwards et al., J. Phys. B, {\textbf{38}}, 363, (2005).
%
\bibitem{anand}
P. Murugandam and S. K. Adhikari, Comp.Phys.Comm. , {\textbf{180}}, 1888, (2009).
%



\end{thebibliography}
\end{document}